\documentclass[aps,prb,twocolumn,superscriptaddress]{revtex4-2}

\usepackage{graphicx}
\usepackage{bm}
\usepackage{wrapfig}
\usepackage[unicode]{hyperref}
\usepackage{amsmath}
\usepackage{amsfonts}
\usepackage{amssymb}

\newcommand{\beq}{\begin{equation}}
\newcommand{\eeq}{\end{equation}}
\newcommand{\beqa}{\begin{eqnarray}}
\newcommand{\eeqa}{\end{eqnarray}}
\newcommand{\eps}{\epsilon}
\newcommand{\rr}{{\bf r}}
\newcommand{\dd}{{\bf \delta}}
\newcommand{\s}{\sigma }
\newcommand{\p}{{\bf p}}

\newcommand{\Q}{{\bf Q}}
\newcommand{\kk}{{\bf k}}

\newcommand{\vv}{{\bf v}}

\begin{document}
\title{Pair density wave solution for self-consistent model}

\author{S. I. Matveenko}
\affiliation{L.D. Landau Institute for Theoretical Physics, RAS, 142432, Chernogolovka, Moscow region, Russia}
\affiliation{Russian Quantum Center, Skolkovo, Moscow 143025, Russia}
\author{ S. I. Mukhin}
\affiliation{National University of Science and Technology MISIS, Moscow 119049, Russia}

\begin{abstract}
In the   self-consistent approximation   for   the  two-dimensional mean field model we found  analytic solution
 for the ground state with coexisting d-wave symmetric bond ordered pair density wave (PDW)
 and spin (SDW) or charge (CDW) density waves, as observed in some high-temperature
 superconductors. In particular, the solution  gives  the same periodicity for CDW and PDW,
 and a pseudogap in the fermi-excitation spectrum.
\end{abstract}

\maketitle

\section{Introduction}
After pioneering work \cite{Aeppli2002} that demonstrated stripe phase inside the cores
 of the Abrikosov's vortices in high-T$_c$ cuprates in magnetic field, new measurements
 discovered even more complicated coexistence patterns \cite{tranq09}. Namely, new
 superconducting states were found with pair density wave (PDW), where momenta
 of the Cooper pairs are nonzero, and the order parameter is nonuniform and oscillatory
 in space.  These states, similar to Fulde-Ferrell-Larkin-Ovchinnikov (FFLO)
 states \cite{FF64, LO65}, can  coexist with spin- or/and charge density waves
 SDW/CDW \cite{Wang18} in the Abrikosov's vortex halo. Moreover, in contrast
 with the  FFLO case, PDW states have been proposed  to exist also in the absence of
 an external magnetic field in a family of cuprate high-temperature superconductors (HTSC),
 where they co-exist with the stripe-phase  \cite{Kato_02, lee18}.  As proposed in \cite{tranq09}, a fluctuating PDW may be responsible for the pseudo-gap structure at the anti-node and the Fermi arc near the node.  The case of fluctuating SDW/CDW with condensed SC densities occupying finite volume in space (Q-balls) was explored for the temperatures above T$_c$, see e.g. \cite{Mukhin(2022), Mukhin(20221), Mukhin(20222)}.  In the latter scenario the non-topological solitons formed by thermodynamic quantum time crystals of SDW/CDW serve as the 'pairing glue' for the formation of Cooper pairs, that condense inside these solitons (called Q-balls) thus lowering the total energy of the Q-balls gas. The idea of Q-balls was first proposed for quark gluon plasma \cite{Coleman}.

Previously, we have presented
 self-consistent solutions in analytic form for the two-dimensional Hubbard t-U-V model
 with $d_{x^2-y^2}$ symmetry of the superconducting PDW order parameter in a weak
 external magnetic field, much less than the first critical field $H_{c1}$, above which
 Abrikosov's vortex would occur \cite{EPL15}. In this state superconducting order changes
 sign when entering the 'stripe-phase' ordered domain, with SDW's envelope forming a single
 stripe. Here we present new self-consistent analytic solution for the ground state with coexisting
 d-wave symmetric bond ordered density waves: PDW, SDW and/or CDW, forming periodic
 stripe-like structure in zero external magnetic field. In the case of bond ordered density waves,
 unlike in the site ordered case considered previously \cite{EPL15}, we find that the pair wave function is intertwined
 with the spin- and charge-stripe order in such a way that the spin order and pair
wave function indeed minimize their overlap, in accord with experimental evidence \cite{Wen}.  Indications of this kind of PDW-SDW-CDW
 pattern were previously found in the Monte-Carlo calculations \cite{Kato_02}. Here we
 demonstrate that  a pseudo-gap like behavior due to periodic structures under doping could be described
 by analytic self-consistent solutions emerging in the vicinity of the hot spots on the Fermi
 surface, with connecting wave vectors serving as the underlying wave vectors of the
 corresponding density waves. This picture is similar to the more simplistic description of the
 periodic 1D CDW/SDW structures (the Peierls instability), where the 'nesting' wave vectors
 depend on doping \cite{Muk, MatMuk, 1d}.

\section{The model }
We start from the two-dimensional mean field Hamiltonian on the square lattice which takes into account
 self-consistently distributions of charge, spin, and superconducting densities, compare with  \cite{lee18}:
\begin{widetext}
\beqa
H = -t\sum_{\langle i,j\rangle,
\sigma}c^{\dagger}_{i,\sigma}c_{j,\sigma} +
\left[\sum_{<i,j>, \s} \Delta_{s}(i,j) \sigma c^{\dagger}_{i,\sigma}
c_{j,\sigma}+
 \sum_{<i,j>, \s} \Delta_{c}(i,j)  c^{\dagger}_{i,\sigma}
c_{j,\sigma}+ \right. \nonumber\\
 \left.\sum_{<i,j>, \s} \Delta_{sc}(i,j;\sigma)c^{\dagger}_{i,\sigma}
c^{\dagger}_{j,-\sigma} + h.c.\right] - \mu \sum_{i, \sigma} \hat{n}_{i, \sigma} ,
  \label{H}
\eeqa
\end{widetext}
where
the first term is the kinetic energy, the next three terms describe spin, charge and
   superconducting correlations, respectively.
A sum $\sum_{<i,j>, \s}$ is taken over nearest neighboring sites
${\bf r}_i$, ${\bf r}_j$ of the square lattice, and spin  components $\s  = \uparrow ,
 \downarrow$. The  spin  and charge density wave terms ($\Delta_s, \, \Delta_c$) in  Eq. (1) are
 written in the bond centered form to take into account possible  $d$-wave symmetric order
 parameters. For the $s$-wave symmetric orders   we previously used the  on-site centered
  terms \cite{MatMuk}:
$$
\sum_{i, \s} \Delta_{s}(i) \sigma c^{\dagger}_{i,\sigma}
c_{i,\sigma}+
 \sum_{i, \s} \Delta_{c}(i) c^{\dagger}_{i,\sigma}
c_{i\sigma}.
$$

The spin ($\Delta_s$), charge ($\Delta_c$) and superconducting ($\Delta_{sc}$) orders satisfy  the self-consistency equations:

\beqa
\Delta^*_s (i,j) =  g_s \langle  \sum_{\sigma}\sigma c^{\dagger}_{i,\sigma}
c_{j,\sigma}  \rangle, \quad \nonumber \\
\Delta^*_c (i, j) = g_c
 \langle  \sum_{\sigma} c^{\dagger}_{i,\sigma}
c_{j,\sigma}  \rangle, \nonumber \\
\Delta_{sc}(i,j;\sigma)= g_{sc} \langle c_{j,-\sigma}
c_{i,\sigma}\rangle,
\label{selfc}
\eeqa
where spin ($g_s$), charge ($g_c$) and superconducting  ($g_{sc}$) coupling  constants depend on the  considered model.
We will further use the system of units such that the period of the lattice equals one, and the energy will be  measured in the units of the nearest neighbour hopping integral $t$.

The Hamiltonian (\ref{H}) has a general form with model-dependent  coupling constants in (\ref{selfc}).   It can be obtained  from the microscopic Hubbard type models  with  the help of Hubbard-Stratonovich transformation of interaction terms.  In the saddle-point approximation, this reduces to the Hamiltonian obtained in the Hartree-Fock approximation. In the case of the t-U-V Hubbard model
\beq
\begin{split}
H =-t\sum_{\langle i,j\rangle,\sigma}c^{\dagger}_{i,\sigma}c_{j,\sigma} + U\sum_i  c^{\dagger}_{i,\uparrow}
c_{i,\uparrow} c^{\dagger}_{i,\downarrow} c_{i,\downarrow}+ \\
V  c^{\dagger}_{i,\sigma} c^{\dagger}_{j,-\sigma} c_{j,-\sigma} c_{i,\sigma}  - \mu \sum_{i, \sigma} \hat{n}_{i, \sigma} ,
 \end{split}
  \label{Hub}
  \eeq
  where $U >0$ is the Hubbard repulsion potential, and $V <0$ is the nearest neighbour interaction potential,
  the  transformation   (see, for example \cite{shulz,MatMuk}) gives Eqs. (1) - (2) with  coupling constants presented in Table \ref{tb}
 
  \begin{table}[h!]
  \begin{center}
 \begin{tabular}{|c|| c | c | c|} 
 \hline
  \, & $g_s$ & $g_c$ & $g_{sc}$ \\
 \hline\hline
 site-type order & $-U/2$ & $U/2$ & $U/2$ \\ 
 \hline
 bond-type order & V & $-V$ & $V$ \\ 
 \hline
\end{tabular}
\end{center}
  \caption{Coupling constants fot the t-U-V model, }
\label{tb}
\end{table}
  
  The  coupling constants for site-type ($i=j$ in Eq.~(2)), order parameters are given in the first row of the table, and for bond-type (where indices $i$ and $j$ numerate  the nearest neighbouring sites) are presented in the second row.
  
  In the pure  Hubbard model ($V=0$)  the superconducting coupling constant $g_{sc} =0$ or arises in the second order with respect to other interaction constants. The ground state of the half-filled system is antiferromagnet, the doping leads to a  periodic spin density structure.
If we add attraction $V<0$ between particles on the nearest neighbouring sites, the ground state becomes superconducting with d-wave symmetry. The ground state with s-wave pairing is impossible due to a strong on-site repulsion ($U>0$).  The charge density wave
structure can appear either in the presence of SDW in the second order of the perturbation theory (with period equal to one-half of SDW period), or as the  main structure in the case of particle attraction, for example.

 The ground state of the system is found  by minimization of the thermodynamic potential.
  The analysis of the system is strongly simplified if we have only two nonzero parameters, for
 example,  $\Delta_s $ and $\Delta_{sc}$, or $\Delta_c$ and $\Delta_{sc}$. In particular,
  in the case of
$\Delta_c (i, j) = \Delta$, if  $ {i = j \pm \hat{y}}$,  and $\Delta_c (i, j) =-\Delta$,  if
 ${i = j \pm \hat{x}}$, we will obtain d-wave symmetric CDW, with the same symmetry as  for superconducting order parameter. Here vectors $\hat{x}, \hat{y}$ connect nearest neighbouring points of the lattice along the $x$ and $y$ axes respectively. Then, Fourier transformed order parameter: $\Delta_{c}(\p,\rr )=\sum_{{\rr}^{\prime}}\Delta_{c} (\rr,{\rr}^{\prime})\exp\{-i\p\cdot(\rr-{\rr}^{\prime})\}\equiv\Delta_\p(\rr) \propto (\cos p_x - \cos p_y )$, and  the periods of CDW and  PDW  will coincide. Here and everywhere below we assume the lattice constants  equal to unity.

  We should emphasize that our effective model approximation, as all mean field type approximations, strictly speaking, is not applicable to the low dimensional case. Considering 2D case, we bear in mind, that real materials are 3D and consist of many layers, so that interlayer   interactions stabilize low dimensional diverging fluctuations and allow to use the mean field approach. Therefore, the considered self-consistent approximation seems justified   to describe some properties of  quasi-two-dimensional compounds.  Of course, properties of pure 2D models  and considered  effective models will be different.

 \subsection{Spin Density Wave and  Superconductivity}

Consider the case of coexistence of SDW and PDW,  as observed in cuprates that are constituted
 by Sr/Ba doped La$_2$CuO$_4$ \cite{Wen, Tsvelik}:

We put $\Delta_c  \to 0$  in  the  Hamiltonian (\ref{H})  and diagonalize it  with the help of
  the Bogoliubov transformations:
\beq
\hat{c}_{\sigma}(\rr) = \sum_{\nu} \gamma_{\nu, \sigma} u_{\nu, \sigma}(\rr)-\sigma
\gamma^+_{\nu, -\sigma} v^*_{\nu, -\sigma}(\rr),
\label{tr1}
\eeq
\beq
\hat{c}^{\dagger}_{\sigma}(\rr) = \sum_{\nu} \gamma^{\dagger}_{\nu, \sigma} u^*_{\nu, \sigma}(\rr)-\sigma
\gamma_{\nu, -\sigma} v_{\nu, -\sigma}(\rr),
\label{tr2}
\eeq
with   new fermion operators satifying the fermion commutation relations $\gamma_{\nu, \sigma}$, $\gamma^+_{\nu, \sigma}$, $\nu = 1, 2,\ldots$.  The Hamiltonian becomes
\beq
H =  E_g + \sum_{\eps_{\nu} >0} \eps_{\nu,\sigma} \gamma_{\nu,\sigma}^+ \gamma_{\nu,\sigma},
\label{eg}
\eeq

where $E_g$ is the  ground state energy and $\eps_{\nu}$ is the energy of excitation $\nu$.  The commutator of $H$ with $\gamma_{\nu, \sigma}$ and  $ \gamma_{\nu,\sigma}^+$  reads
\beq
[H,  \gamma_{\nu,\sigma}] = - \eps_{\nu, \sigma}  \gamma_{\nu,\sigma}, \quad [H,  \gamma^+_{\nu,\sigma}] = \eps_{\nu, \sigma}  \gamma_{\nu,\sigma}^+.
\label{commut}
\eeq
To derive the equations for functions $u$, $v$ we calculate  the commutator
\beqa
&&[\hat{c}_{\sigma}(\rr) , H]=-t\sum_{\rr'} \hat{c}_{\sigma}(\rr') -\mu(\rr) \hat{c}_{\sigma}(\rr)   +
\Delta_s (\rr) \sigma \hat{c}_{\sigma}(\rr) \nonumber \\
& &+\sum_{\rr'} \{ \Delta_{sc}(\rr, \rr';\sigma )\hat{c}_{-\sigma}^{\dagger}(\rr') - \Delta_{sc}(\rr', \rr;-\sigma )\hat{c}_{-\sigma}^{\dagger}(\rr')\},
\label{comm2}
\eeqa

We replace operators $ \hat{c}_{\sigma}(\rr)$, $ \hat{c}_{\sigma}^{\dagger}(\rr)$ by the $\gamma_{\nu, \sigma}$'s by means of (\ref{tr1}), (\ref{tr2}), and apply the commutation relations (\ref{commut}). Comparing  the coefficients
of $\gamma_{\nu, \sigma}$, and $\gamma_{\nu, \sigma}^{\dagger}$ on the two sides of Eq.~(\ref{comm2}),
we obtain the
  eigenvalue equations:
\beqa
-t\sum_{\dd} u_{\s}(\rr +\dd ) - \mu u_{\s}(\rr)&
+\Delta_s(\rr) \s u_{\s}(\rr ) \nonumber \\
+\sum_{\dd} \Delta(\rr,\rr +\dd;\s )\s v_{\s} (\rr +\dd )&
 =\eps_{\s} u_{\s}(\rr),
 \label{deq1}
\eeqa
\beqa
-\sum_{\dd} \Delta^*(\rr,\rr +\dd;-\s )\s u_{\s} (\rr +\dd ) &
+t\sum_{\dd} v_{\s}(\rr +\dd ) \nonumber \\
+ \mu v_{\s}(\rr)
+\Delta_s (\rr) \s v_{\s}(\rr ) & = \eps_{\s} v_{\s}(\rr),
\label{deq2}
\eeqa
where $\dd=\pm \hat{\bf x}, \pm \hat{\bf y}$ and $ \Delta$ is short notation for
 superconducting order $\Delta_{sc}$ introduced in Eq.~(\ref{selfc}).

We suppose the $d_{x^2 -y^2}$ symmetry of the
 superconducting order parameter $\Delta_{sc}(\rr ,\rr \pm \hat{\bf x};\s)=
 \s \Delta_d (\rr )$, $\Delta_{sc}(\rr ,\rr \pm \hat{\bf y};\s)=
 -\s \Delta_d (\rr )$ so that the Fourier transform has the form $\Delta_{sc} (\rr, \p)=\sum_{{\rr}^{\prime}}\Delta_{sc} (\rr,{\rr}^{\prime})\exp\{-i\p\cdot(\rr-{\rr}^{\prime})\}
 \equiv \Delta_{\p} (\rr ) = 2(\cos p_x - \cos p_y ) \Delta_d (\rr)$ , with slowly varying function $\Delta_d(\rr )$.
The system (\ref{deq1}) -- (\ref{deq2}) can be rewritten in the continuous  approximation.

\begin{figure}
 \centering
    \includegraphics[width=0.8\linewidth]{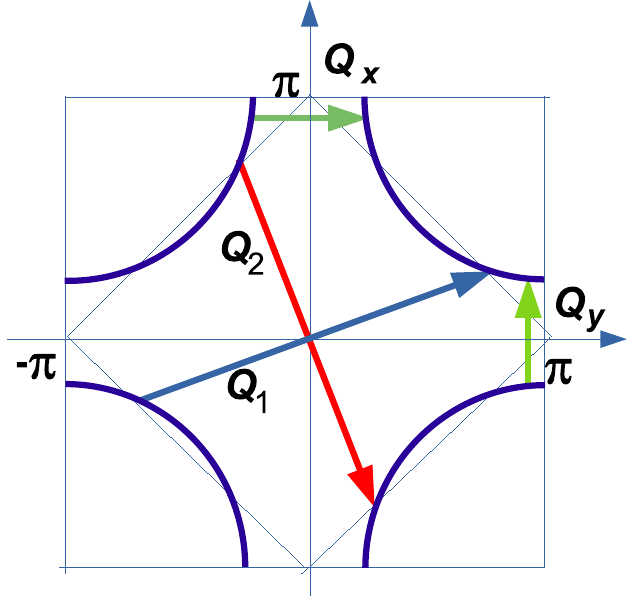}
\caption{The Fermi surface in the Brillouin zone for a nearly half-filled square lattice model
 and vectors $\Q_{1,2}$ and $\Q_{x,y}$ of SDW, CDW, connecting the 'hot spots' of the Fermi surface. }
\label{fs}
\end{figure}

Consider states near the Fermi surface (FS)  (see Fig.1) and use linear approximation for the
 quasiparticles spectrum.
 We write the functions $u(\rr )$ and $v(\rr )$ as:
\beq
u_{\s}(\rr ) =\sum_{n}\sum_{\p \in FS}[u_{\p, \s}(\rr ) e^{i \p \rr}
+ \s u_{\p - \Q_n, \s} (\rr ) e^{i(\p -\Q_n )\rr }],
\label{bdg-spin}
\eeq
\beq
v_{\s}(\rr ) =\sum_{n}\sum_{\p \in FS}[v_{\p, \s}(\rr ) e^{i \p \rr}
+ \s v_{\p - \Q_n, \s} (\rr ) e^{i(\p -\Q_n )\rr }],
\label{bdg-spin2}
\eeq
where $n=1,2$, and   $\Q_1$, $\Q_2$ are vectors of a bidirectional SDW.
For a small doping these vectors are close to the antiferromagnetic ones: $ \Q_1 = (\pi, \pi) + \delta \Q $,  $|\Q_1| = |\Q_2|$, $\Q_1 \perp \Q_2$.
In the general case of a  doped system vectors $\Q_{n}$ are incommensurate with reciprocal lattice vectors.
Note, that vectors $\Q_n$  connect $\p, \p -\Q_n$-states from the d-wave segments with
 the same sign (as is opposite to the case considered previously in \cite{EPL15}).

 We rewrite
  the SDW  order parameters  as
\beq
\Delta_s(\rr) =  \sum_{n} \Delta_{s,n}  (\rr )\exp (i \Q_n
\rr )  + h.c.,\label{SDW}
\eeq
with slowly varying functions $\Delta_{s,n }(\rr )$.
Eigenvalue equations (\ref{deq1}), (\ref{deq2})
 take the form $\hat{H}\Psi = E \Psi$, with the Hamiltonian operator:
 \begin{widetext}
\beq
  \hat{H} =
\begin{pmatrix}
-i\vv_{\p}\nabla_{\rr} +\eps_{\p} -\mu&
 \Delta_{s,n} (\rr) & \Delta_{-\p} &0 \\
\Delta^*_{s,n} (\rr)  & -i\vv_{\p-\Q}\nabla_{\rr} +\eps_{\p -\Q} -\mu
 &0& \Delta_{-(\p -\Q)}\\
 \Delta_{\p}^*  & 0 & i\vv_{\p}\nabla_{\rr} -\eps_{\p}+
\mu & \Delta_{s,n} (\rr) \\
 0& \Delta_{\p -\Q}^* & \Delta^*_{s,n}(\rr ) & i\vv_{\p
-\Q}\nabla_{\rr}
 -\eps_{\p -\Q}+ \mu
\end{pmatrix},
\label{HH}
\eeq
\end{widetext}
where we have omitted the lower index $sc$ in the notation of the superconducting order parameter $\Delta_{sc} (\rr, \p)\rightarrow \Delta_{\p}$ defined above, and $\eps_{\p} = -2t(\cos p_x + \cos p_y ) -\mu$,
${\bf v}_{\p}=2 t (\sin p_x, \sin p_y )$.

We have linearized free  particle spectrum  near  the Fermi surface  (FS) in the Eq.(\ref{HH}).
 Note, that at zero temperature we have at the FS the identity $\eps(\p) = \eps (\p - \Q )= \mu$.
For the case of $d_{x^2 -y^2}$ symmetry we obtain  $\Delta_{-\p}=
\Delta_{\p}=\Delta_{\p -\Q}= 2(\cos p_x - \cos p_y ) \Delta_d
(\rr)$, since vectors $\p$ and $\p - \Q$ are symmetric  either with  respect to the origin
 point $\p=0$, then ($-\p = \p - \Q_{\pm}$), or with  respect to the axis $p_x$ or $p_y$ ($-\p_{x,y} = \p - \Q_{x,y}$), see Fig. 1.

 For the case of site-type SDW we  consider states with d-wave symmetry of the SDW order parameter in the same way as
 for described  above  SC  case. As a result we simply
add the index $\p$ to the  SDW order parameter  ($ \Delta_s \to \Delta_{s,\p}$) in Eq.~(\ref{HH}).

In the case of  coordinate-independent amplitudes in Eq. (\ref{SDW}) and analogously for PDW wave : $\Delta_{s,\p}(r) = \Delta_{s, \p} , \, \Delta_{\p}(r) = \Delta_{\p}$, in the CDW - free case, the eigenvalue spectrum has the form:

\beq
E^2(\p) = \xi_{\p}^2 + (\Delta_{\p}\pm \Delta_{s,\p})^2,
\label{spec}
\eeq
where $\xi_{\p}= \vv_{\p_F}\centerdot (\p -\p_F)$, and $\p_F  \in  FS$.
The gapped
spectra in Eq. (\ref{spec}) characterize only momenta $p$ in the
vicinities of the hot spots on the Fermi surface connected
by the wave vectors $\Q_i$, that play the role of the wave
vectors of the unidirectional density waves considered
here. Hence, obtained spectra have a pseudogap structure.
Moreover, we obtain the pseudogap spectra every
time when SC coexists with SDW or CDW. Besides, the presence of $\pm$ sign in Eq. (\ref{spec}) signifies two Bogoliubov bands with smaller $|\Delta_{\p}- \Delta_{s,\p}|$ and greater $|\Delta_{\p}+ \Delta_{s,\p}|$ gaps at the Fermi level. This, in principle, may cause a zero-bias anomaly \cite{Yuli} of  the tunneling current along the c axis perpendicular to the $a-b$ plane, when the tunneling took place out of the antinodal direction.

In the general case a solution of the system of equations (\ref{HH}) is unknown.
But for quasi-1D structures we can use  the ansatz  which was applied for 1D  model
\cite{EPL15,1d}
\[
v_{\pm} (\rr) = \gamma_{\pm} u_{\mp}(\rr),
\]
with constant $\gamma_{\pm}$.
For the case
\[
\Delta_{s,\p} (\rr ) = \vert \Delta_{s,\p} (\rr) \vert e^{ i \varphi_s},\,
\Delta_{\p} (\rr) = \vert \Delta_{\p} (\rr)\vert e^{i\varphi},\,
\varphi,\, \varphi_s = const,
\]
the ansatz is satisfied at
$
\gamma_+ = \pm e^{i(\varphi -\varphi_s )}, \,
\gamma_- = \pm e^{-i(\varphi + \varphi_s)}$,
 and  4x4  matrix equations  (\ref{HH})  are reduced to 2x2  BdG type system
\beqa
-i\vv_{\p}\nabla u_+ + \tilde{\Delta}_{\p}(\rr)u_- = E u_+\label{se0}\\
\tilde{\Delta}_{\p}^*(\rr)u_+ + i\vv_{\p}\nabla u_- = Eu_-
\label{se}
\eeqa
with function $\tilde{\Delta}_{\p}(\rr) = ( \Delta_{s,\p}(\rr) \pm \Delta_{\p}(\rr))
e^{i\varphi}$.  Equations (\ref{se0}), (\ref{se}) are exact, provided that phases
$\varphi$,  $\varphi_s$ are constant or slowly varying in space functions.

The one-dimensional analogue of these equations
are eigenvalue equations for the Peierls model.   Exact  solutions describing solitons,  polarons and CDW periodic structures
as a function of doping (hole concentration) were studied in details \cite{Peierls1,Peierls2}.
Consider one-stripe structure  (or domain wall) aligned vertically along the $y$-direction (or  horizontally along $x$). 
The  solution is the same as in 1D case:
\beq
\tilde{\Delta}^{\pm}_{\p}(x) = \pm \Delta_{\p}  \tanh (\Delta_{\p} x/v_{\p}  \pm a/2),
\label{kink}
\eeq
\noindent
where the dimensionless parameter $a$ is found by the minimization of the free energy.
 The nonzero value  $a$ is reached
in the  region $0<|g_s- g_{sc}| \ll g_{sc}$.

In this region we have  nonzero  both superconducting and spin order parameters:
\begin{widetext}
\beq
\Delta_{sc,\p}  =(\tilde{\Delta}^+_p - \tilde{\Delta}^-_p )/2  =
 \Delta_{\p} \tanh (\Delta_{\p} x/v_{\p}) \frac{\cosh^2(\Delta_{\p} x/v_{\p})}{\cosh^2(\Delta_{\p} x/v_{\p}) + \sinh^2(a/2)}
 \label{deltas}
\eeq
\beq
 \Delta_{s,\p}= (\tilde{\Delta}^+_p + \tilde{\Delta}^-_p )/2  =
         \Delta_p \tanh (a/2) \frac{\cosh^2(a/2)}{\cosh^2(a/2) + \sinh^2(\Delta_{\p} x/v_{\p})}
\label{deltasc}
\eeq
\end{widetext}
At finite doping concentrations a periodic structure (PDW + SDW) arises with  the solution:
\beqa
\tilde{\Delta}^{\pm}_{\p} = \pm\Delta_{\p} \tanh (\Delta_{\p} x/v_{\p}  \pm a/2) \to \nonumber \\
\pm\Delta_{\p}\sqrt{k}\, \rm{sn}(\Delta_{\p}x/(v_{\p}\sqrt{k})  \pm a/2, k)  \nonumber \\
 \to \Delta_{\p}\sin(2\pi x/l \pm a/2),
\label{periodic}
\eeqa
where $ \rm{sn} (\Delta_{\p}x/(v_{\p}\sqrt{k}) , k )$ is the Jakobi elliptic function
with the parameter $0<k<1$ defined by the period of the structure
 $ l = 4 \pi K(k)\sqrt{k}v_{\p}/\Delta_{\p} $ ( $l = 2 /|\rho -1|$ for purely 1D model),
 where $K(k)$ is the complete elliptic integral
 of the first kind. Parameter $k$ varies from $k=1$  where
$ {\rm sn}(\Delta_{\p}x/(v_{\p}\sqrt{k}), k) = \tanh (\Delta_{\p} x/v_{\p})$, to $k \ll 1$ where
${\rm sn} (\Delta_{\p}x/(v_{\p}\sqrt{k}), k) \sim \sin(2 \pi x /l)$.

\begin{figure}
  \centering
  \includegraphics[width=0.8\linewidth]{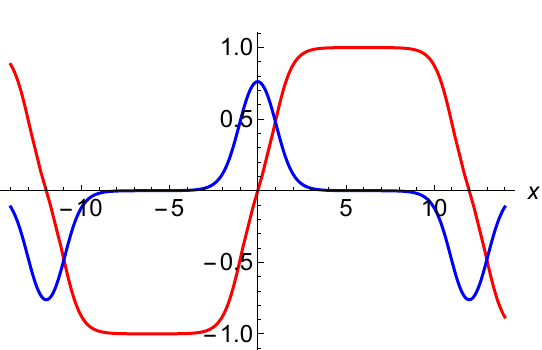}\\
  \caption{PDW order parameter $\Delta_{sc} (x)$ (red) and SDW  order parameter  $\Delta_s(x)$ (blue) in dimensionless units as a function of dimensionless coordinate $x$ (one period of superstructure is depicted).  }
  \label{sdw-pdw}
\end{figure}

\begin{figure}
 \includegraphics[width=0.8\linewidth]{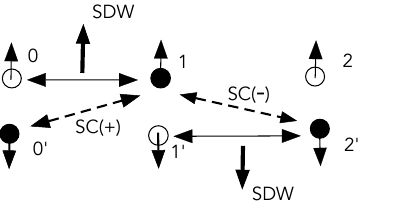}\\
  \caption{Permutation of the spin-up and spin-down states along $0,1$ and $1,2$ bonds in the e.g. $\{0',1\}$ and $\{1,2'\}$ Cooper pairs of the PDW order parameter  is caused by the corresponding change of sign within the antiferromagnetic SDW order on the same nearest neighbouring bonds. }
  \label{prs}
\end{figure}

The solutions (\ref{deltas}), (\ref{deltasc}) have the form of a sum or difference of two soliton (kink) solutions (\ref{kink}), with distance between them proportional to the  dimensionless parameter  $a$.  The value of $a$ is found from the
minimization of the total energy. Two phases are coexisting   only when $a \neq 0$.  
 The typical picture of coexisting order parameters is depicted in  Fig.~\ref{sdw-pdw},  where we used values $k = 0.99$ and $a = 1$.
We see in Fig. \ref{sdw-pdw} that PDW superconducting order changes sign inside  alternating domains divided by the SDW or CDW domains, see Fig. \ref{p-c}, in the unidirectional structures derived in Eq. (\ref{periodic}), and Eq. (\ref{pdw-cdw}) below. 
The change of sign, $\pm SC$, of the SC order parameter follows the change of sign of the spin in the SDW order in Fig. \ref{prs}. The alternating filled and empty circles that designate occupied and unoccupied sites in Fig. \ref{prs} cause in an obvious way the sign change of the PDW bond order parameter defined in the third line of Eq. (\ref{selfc}) due to the obvious permutation of the  $c_{i,\sigma}$ and  $c_{j,-\sigma}$  fermionic operators in the definition of the bond superconducting order on the neighbouring bonds $0,1$ and $1,2$.

This behaviour was recently inferred from STM experimental data in LBCO compounds \cite{Wang} where tunneling current along the c axis has revealed a zero-bias anomaly \cite{Yuli}.

We did not examine here a charge density distribution (we put $\Delta_c  \to 0$).  But  we can  investigate it with the help of
a perturbation theory.   It is easy to see that  the SDW
 generates  (in the second order  in $\Delta_s$)  the CDW with a small  amplitude
$\Delta_{CDW} (\rr) \propto\int D\Psi D{\Psi^+}\Psi^+_{p,\sigma} \Psi_{p-Q, \sigma} e^{-S_0 -\int \Delta_{SDW} \Psi^+_{p,\sigma'} \Psi_{p-Q, -\sigma'}}$. Expanding the exponent into a series in $\Delta_{SDW}$ we obtain that  $\Delta_{CDW}(\rr) \propto \Delta_{SDW}^2(\rr)$.  Therefore wave vectors of CDW and SDW structures are related as   $2\Q_{CDW}= \Q_{SDW}$.

In our approach the symmetry of  SDW/CDW is defined by the symmetry
of PDW. That is, if PDW has the d-wave symmetry (there is no s-wave ground state due to large on-site repulsion)  then induced SDW/CDW will have d-wave symmetry.

\subsection{Charge Density Wave and  Superconductivity}

Now, consider the case of coexistence of CDW and PDW,  as observed in e.g. YBCO doped
 compounds \cite{Frad16,cdw-ybacuo}.

Eigenvalue equations differ from (\ref{deq1})- (\ref{deq2}) by substitution
$\s \Delta_s  \to \Delta_c$.
In the general case  we rewrite  the CDW order parameter  as
\[
\Delta_c(\rr) =  \Delta_{c}  (\rr )\exp (i \Q
\rr )  + h.c.
\]
with slowly varying function $\Delta_{c}$.
Consider the experimentally observed case of the CDW  wave vector along the horizontal
 axis: $\Q = \Q_x $, as shown in Fig.1.
The general form of the superconducting order parameter  reads:
\[
 \Delta_{\p} (\rr) =\Delta_{\p, 1}(\rr)  + \Delta_{\p, 2} \exp i\Q_x \rr + h.c.,
\]
where the first term is the contribution from the usual pairing with zero total momentum of pairs, and   the second oscillating term describes
 PDW. It  occurs due to pairing of particles with nonzero total momenta $-\kk_F+\Q/2, +\kk_F + \Q/2$.

Instead of transformations  (\ref{bdg-spin}) -  (\ref{bdg-spin2}) we use  the same ones, but
 without the multiplier $\s$ in the second term:
\beq
u_{\s}(\rr ) =\sum_{\p \in FS, p_x >0}[u_{\p, \s} e^{i \p \rr}
+  u_{\p - \Q, \s} (\rr ) e^{i(\p -\Q )\rr }],
\eeq
We obtain instead of (\ref{HH}) the  eigenvalue equations:
 $\hat{H}\Psi = E \Psi$, with the Hamiltonian operator:
 \begin{widetext}
\beq
  \hat{H} =
\begin{pmatrix}
-i\vv_{\p}\nabla_{\rr} +\eps_{\p} -\mu&
 \Delta_{c} (\rr) & \Delta_{-\p,1} &\Delta_{-\p,2} \\
\Delta_{c}^* (\rr)  & -i\vv_{\p-\Q}\nabla_{\rr} +\eps_{\p -\Q} -\mu
 & \Delta_{-(\p -\Q),2}& \Delta_{-(\p -\Q),1}\\
 \Delta_{\p,1}^*  & \Delta_{\p,2}^* & i\vv_{\p}\nabla_{\rr} -\eps_{\p}+
\mu & \Delta_{c} (\rr) \\
  \Delta_{\p -\Q,2}^* & \Delta_{\p -\Q,1}^* & \Delta_{c}^*(\rr ) & i\vv_{\p
-\Q}\nabla_{\rr}
 -\eps_{\p -\Q}+ \mu
 \end{pmatrix}.
\label{HH2}
\eeq
\end{widetext}
Note, that for the case $\Delta_{\p, 2} =0$  the Hamiltonian is equivalent to the one obtained for the case of a spin density wave (\ref{HH}).

 Consider another interesting case  of  CDW combined with pure PDW ($ \langle \Delta_p (\rr) \rangle =\Delta_{p, 1}=0$).
Similar to the previous section, we obtain instead of (\ref{se0}), (\ref{se})   the following
effective  2x2 equations
\beqa
-i\vv_{\p}\nabla u_+ + {\Delta_c}(\rr)u_- = (E \mp |\Delta_p| )  u_+\label{se-1}\\
{\Delta_c}^*(\rr)u_+ + i\vv_{\p}\nabla u_- = (E \mp |\Delta_p| ) u_-,
\label{se-2}
\eeqa
where we used the symmetry of the order parameter: $ \Delta_{\p - \Q_x}= \Delta_{\p}$, since
 $ \p = \{p_x, p_y\}$, $\p - \Q_x = \{-p_x, p_y\}$.

Note, that again in the case of  coordinate-independent amplitudes of CDW and PDW waves: $\Delta_{c}(r) = \Delta_{c} , \, \Delta_{\p}(r) = \Delta_{\p}$, solution has the two-branch excitation spectrum:
\beq
E(\p) = \vert \sqrt{ \xi_p^2 + |\Delta_c|^2} \pm |\Delta_{\p}| \vert
\label{esc}
\eeq
(For the case of $d$-wave symmetry of the CDW order parameter  we should substitute
$\Delta_c \to  \Delta_{c,\p} =  (\cos p_x -\cos p_y)\Delta_c$.)

\begin{figure}
  \centering
  \includegraphics[width=0.8\linewidth]{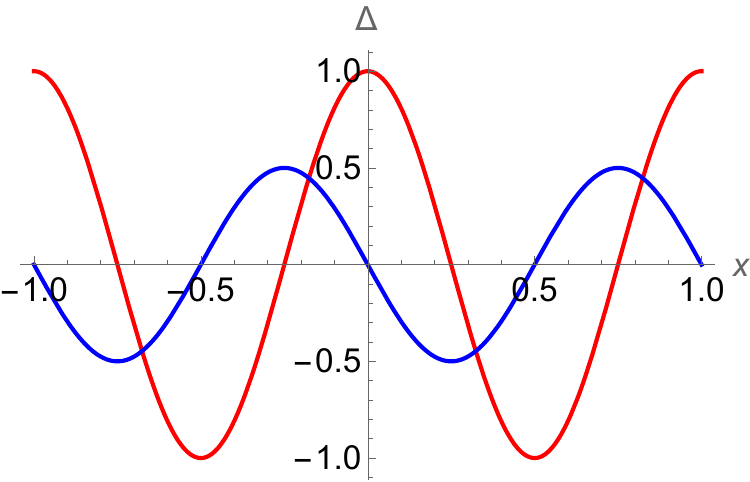}\\
  \caption{PDW (red) and CDW (blue) order parameters.}
  \label{p-c}
\end{figure}

This solution describes  coexisting   CDW  and PDW,  both with the same wave vector
 along the horizontal axis $\hat{x}$ and both having the same period $=2\pi/Q_x$,
 as is observed e.g. in the field induced PDW state in the halo surrounding the vortex core
 in Bi$_2$Sr$_2$CaCu2O$_8$ \cite{pdwH18}:
\beq
  \rho(x) -\bar{\rho} \sim \Delta_c  \cos (Q_x x + \phi),
 \Delta (x) -\bar{\Delta} \sim \Delta_p \cos (Q_x x),
\label{pdw-cdw}
\eeq
where the phase  between PDW and CDW is defined  from the minimization of the total energy. The case $\phi = \pm\pi/2$
corresponds to the competition of CDW and PDW \cite{cdw16,chubuk151}, when zero value of SC order parameter and the maximum of CDW   density amplitude are
reached at the same point, as shown in Fig. \ref{p-c}.

\subsection{Conditions of coexistence SDW/CDW with superconductivity}

We will show  that the PDW phase can coexist with SDW (for certainty), even in the zero magnetic field, in some range of parameters.
The total energy functional has the form:
\beq
W= \sum E(k)     +\int d\rr \left\{ \frac{|\Delta_s|^2}{|g_s|} +\frac{|\Delta_{sc}|^2}{|g_{sc}|}+ \alpha \frac{ \rho(\rr )^2}{2} \right\},
\eeq
where $\Delta_s$, $\Delta_{sc}$ are given by Eqns. (\ref{deltas}), (\ref{deltasc}),  and the parameter $\alpha$ is equal to:
   $\alpha =2 |V|- U/2$, for the t-U-V model.  If we do not specify  the microscopic model, then all coupling constants can take arbitrary values.    For $a=0$ these equations describe
one  kink (domain wall, soliton) in antiferromagnetic (AFM) phase, where superconductivity (SC) appears  as a result of  (hole) doping. The further doping leads to a periodically modulated spin  structure (SDW). For  $a \neq 0$  a periodic SC structure (PDW)  is also formed.

\begin{figure}
 \centering
    \includegraphics[width=0.8\linewidth]{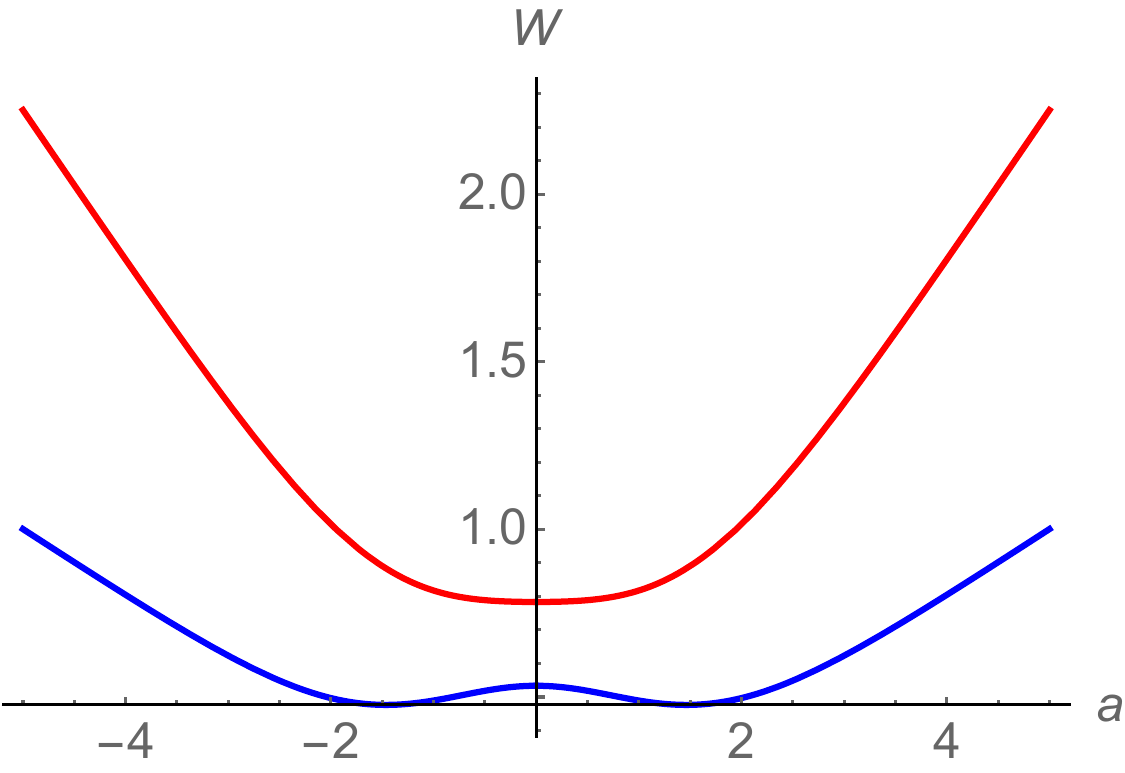}
\caption{The energy $W(a)$  (blue curve) has minimum at $a \neq 0$  for parameters satisfying (\ref{llsc}),  otherwise $a_{min} =0$ (red curve). }
\label{w(a)}
\end{figure}

The local electric charge $ \rho = [u_+^* u_+ + u_-^* u_-] $  equals to:
\beq
 \rho \propto \frac{1}{\cosh^2 (\Delta_{\p} x/v_{\p}+ a/2)} +  \frac{1}{\cosh^2 (\Delta_{\p} x/v_{\p} - a/2)},
\eeq
Since the excitation energies are independent of  the parameter $a$,  only the potential energy defines the minimum with respect to $a$.
As a result the total energy acquires the form:
\beq
W(a) - W(0)  \propto \left\vert \frac{1}{|g_s|} - \frac{1}{|g_{sc}|}\right\vert \frac{2 a}{ \tanh a} + \frac{ 4\alpha}{\sinh^2 a}\left(\frac{a}{\tanh a} -1\right).
\eeq
There is a nontrivial minimum ($a_{min} \neq 0$) in the region where coupling constants $g_s$, $g_{sc}$ are close to each other:
\beq
0 < \left\vert \frac{1}{|g_s|} - \frac{1}{|g_{sc}|}\right\vert < 0.8  \alpha,
\label{llsc}
\eeq
provided that $\alpha >0$.
In this region both SDW and PDW phases exist. 
Recall that the values $g_s$,  $g_{sc}$ and $\alpha$ here are dimensionless.
 A typical behavior of the energy $W(a)$  is shown in Fig.~\ref{w(a)}. 

Note that this effect of coexistence of SDW and PDW  takes place at zero magnetic field, provided (\ref{llsc}) is fulfilled.  If the  condition  (\ref{llsc}) is not  valid then the external magnetic field is necessary  to  stabilize SDW phase.

For the case of the t-U-V model we have  the  conditions $g_s = g_{sc}$ and $a_{min} = 0$. The nontrivial minimum and coexistence of SDW and PDW phase will appear  if we go beyond the mean field  approximation. 

\section{Conclusions}
 Based on  a simple 2D t-U-V microscopic Hubbard model on a square lattice we found different  solutions in analytic form, describing  periodic charge-spin  and  superconducting pair density structures, that coexist in zero external magnetic field. We have defined conditions where coupling constants $g_{sc}$, and $g_s $ (or $g_c$) are close to each other for these solutions to exist. Though so far the derivation is made at zero temperature, nevertheless, this result resonates with the recent proposal \cite{Wang} that PDW is the "mother state" forming anti-nodal gap in the pseudo-gap state above T$_c$ in high-T$_c$ cuprates, that turns via Josephson couplings into bulk superconducting state below T$_c$.  Locally static charge/spin  and superconducting states compete with each other: a decrease in one parameter induces an increase in another.
This is a possible reason for the appearance of pair density waves:  space oscillations of SDW/CDW generate  space oscillations of the superconducting order parameter in such a way that maximum of  PDW density appears in the regions where SDW/CDW vanishes by changing sign and vice versa for SDW/CDW.
On the other hand, in external magnetic field a decrease in the  bulk superconducting  order parameter  in  a vortex region  results in an appearance of  SDW/CDW waves \cite{Aeppli2002}. We have also derived analytical expressions for the fermionic band structure of superconductors with co-existing PDW and CDW/SDW orders, see Eqs. (\ref{spec}) and (\ref{esc}). These solutions posses Bogoliubov double-band structure with smaller and greater gaps at the Fermi level. This, in principle, may cause a zero-bias anomaly \cite{Yuli} of  the tunneling current along the c axis perpendicular to the $a-b$ plane, when the tunneling takes place out of the antinodal direction.

\section{Acknowledgements}
 The work of S.I.M. was in part supported by the universities leadership program "Priority 2030"  in the framework of  NUST "MISIS" grant No. K2-2022-025.

  \end{document}